\documentclass{mn2e}
\usepackage{epsfig}
\usepackage{epsf}
\usepackage{graphicx}
\title[SN~2010jp]{SN~2010jp (PTF10aaxi): A Jet-Driven Type~II Supernova}

\author[Smith et al.]{Nathan Smith$^1$\thanks{Email:
    nathans@as.arizona.edu}, S. Bradley Cenko$^2$, Nat Butler$^2$,
  Joshua S.\ Bloom$^2$, \newauthor Mansi M.\ Kasliwal$^3$, Assaf
  Horesh$^3$, Shrinivas R.\ Kulkarni$^3$, Nicholas M. Law$^4$,
  \newauthor Peter E. Nugent$^{2,5}$, Eran O. Ofek$^{3,6}$, Dovi
  Poznanski$^{2,5,6}$, Robert M. Quimby$^3$, \newauthor Branimir
  Sesar$^3$, Sagi Ben-Ami$^7$, Iair Arcavi$^7$, Avishay Gal-Yam$^7$,
  David Polishook$^7$, \newauthor Dong Xu$^7$, Ofer Yaron$^7$,
  Dale A. Frail$^8$, \& Mark Sullivan$^9$  \\
  $^1$Steward Observatory, University of Arizona, 933 North Cherry
  Avenue, Tucson, AZ 85721, USA \\
  $^2$Department of Astronomy, University of California, Berkeley, CA
  94720-3411, USA \\
  $^3$Cahill Center for Astrophysics, California Institute of
  Technology, Pasadena, CA, 91125, USA \\
  $^4$Dunlap Institute for Astronomy and Astrophysics, University of
  Toronto, 50 St. George Street, Toronto M5S 3H4, Ontario, Canada \\
  $^5$Computational Cosmology Center, Lawrence Berkeley National
  Laboratory, 1 Cyclotron Road, Berkeley, CA 94720, USA \\
  $^8$Einstein Fellow \\
  $^7$The Weizmann Institute of Science, Rehovot 76100, Israel \\
  $^8$National Radio Astronomy Observatory, P.O. Box O, Socorro, NM
  87801, USA \\
  $^9$Department of Physics (Astrophysics), University of Oxford, 
  Keble Road, Oxford, OX13RH, UK}

\begin{document}
\date{Accepted 0000, Received 0000, in original form 0000}
\pagerange{\pageref{firstpage}--\pageref{lastpage}} \pubyear{2002}
\def\arcdeg{\degr}
\maketitle
\label{firstpage}

\begin{abstract}

  We present photometry and spectroscopy of the peculiar Type~II
  supernova (SN) 2010jp, also named PTF10aaxi.  The light curve
  exhibits a linear decline with a relatively low peak absolute
  magnitude of only $-$15.9 (unfiltered), and a low radioactive decay
  luminosity at late times that suggests a low synthesized nickel mass
  of M($^{56}$Ni) \, $\la$ \, 0.003 \, $M_{\odot}$.  Spectra of
  SN~2010jp display an unprecedented {\it triple-peaked} H$\alpha$
  line profile, showing: (1) a narrow (FWHM $\ga$ 800 km s$^{-1}$)
  central component that suggests shock interaction with a dense
  circumstellar medium (CSM); (2) high-velocity blue and red emission
  features centered at $-$12,600 and $+$15,400 km s$^{-1}$; and (3)
  very broad wings extending from $-$22,000 to $+$25,000 km s$^{-1}$.
  These features persist over multiple epochs during the $\sim$100
  days after explosion.  We propose that this line profile indicates a
  bipolar jet-driven explosion, with the central component produced by
  normal SN ejecta and CSM interaction at mid and low latitudes, while
  the high-velocity bumps and broad line wings arise in a
  nonrelativistic bipolar jet.  Two variations of the jet
  interpretation seem plausible: (1) A fast jet mixes $^{56}$Ni to
  high velocities in polar zones of the H-rich envelope, or (2) the
  reverse shock in the jet produces blue and red bumps in Balmer lines
  when a jet interacts with dense CSM.  Jet-driven SNe~II are
  predicted for collapsars resulting from a wide range of initial
  masses above 25 $M_{\odot}$, especially at sub-solar metallicity.
  This seems consistent with the SN host environent, which is either
  an extremely low-luminosity dwarf galaxy or the very remote parts of
  an interacting pair of star-forming galaxies.  It also seems
  consistent with the apparently low $^{56}$Ni mass that may accompany
  black hole formation.  We speculate that the jet survives to produce
  observable signatures because the star's H envelope was very low
  mass, having been mostly stripped away by the previous eruptive mass
  loss indicated by the Type IIn features in the spectrum.

\end{abstract}


\begin{figure*}\begin{center}
\includegraphics[width=7.1in]{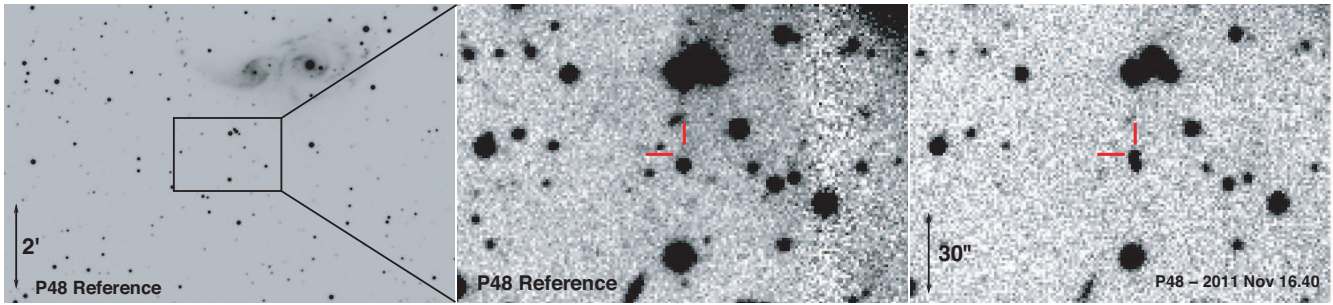}
\end{center}
\caption{Finder chart for SN\,2010jp (PTF10aaxi).  \textit{Left:}
  Wide-field P48 reference image of the SN field, including
  pre-discovery frames taken from 2009 November to 2010 March.  The
  interacting galaxy pair IC\,2163 / NGC\,2207, at the same redshift
  as SN\,2010jp, is clearly visible about 2 arcminutes to the north.
  \textit{Center:} Zoomed-in portion of the same P48 reference image
  of the location of SN\,2010jp.  The SN position is indicated with
  the red tick marks (although the SN itself is not in this image).
  \textit{Right:} P48 discovery image. The SN position, indicated
  again with the red tick marks, is only a few arcseconds north of a
  foreground star, which appears blended with the relatively coarse
  P48 angular resolution.}\label{fig:img}
\end{figure*}

\section{INTRODUCTION}

Many theoretical studies of the core-collapse mechanism suggest that
breaking spherical symmetry may be an essential ingredient in
overcoming a stalled shock and producing a successful supernova (SN)
explosion (Blondin et al.\ 2003; Buras et al.\ 2006a, 2006b; Burrows
et al.\ 2006, 2007).  An extreme case of breaking spherical symmetry
involves jet-driven explosions (Khokhlov et al.\ 1999; H\"oflich et
al.\ 2001; Maeda \& Nomoto 2003; Wheeler et al.\ 2000; Couch et al.\
2009).  Strongly collimated jets that expel the surrounding stellar
envelopes may arise from accretion onto newly-formed black holes as in
the ``collapsar'' model (MacFadyen \& Woosley 1999; MacFadyen et al.\
2001), or by magnetohydrodynamic (MHD) mechanisms in the collapse and
spin-down of highly magnetized and rapidly rotating neutron stars, or
magnetars (LeBlanc \& Wilson 1970; Bodenheimer \& Ostriker 1974;
Wheeler et al.\ 2000; Thompson et al.\ 2004; Bucciantini et al.\ 2006;
Burrows et al.\ 2007; Komissarov \& Barkov 2007; Dessart et al.\ 2008;
Metzger et al.\ 2010; Piro \& Ott 2011). A jet-driven explosion may be
particularly important for producing successful SNe from high-mass
stars above $\sim$25 $M_{\odot}$ (e.g., MacFadyen et al.\ 2001; Heger
et al.\ 2003), which might otherwise collapse quietly to a black hole.

The clearest observational evidence for jets in core-collapse SNe
comes from the association of broad-lined SNe~Ic with the relativistic
jets that produce observable gamma-ray bursts (GRBs) (Woosley \& Bloom
2006; Galama et al.\ 1998; Matheson et al.\ 2003; Mazzali et al.\
2005).  Launched from deep within the collapsing stellar core, the
escape of the jet is facilitated by the fact that the progenitor star
was compact and had all of its H envelope (and most or all of its He
envelope) stripped away during its prior evolution.  Jets are expected
to not survive passage through the larger and more massive H envelope
of a typical RSG, because jet kinetic energy gets thermalized by the
large H envelope (MacFadyen et al.\ 2001; H\"oflich et al.\ 2001).  In
some cases this scenario may yield a relatively weak explosion powered
by accretion onto the black hole (e.g., Quataert \& Kasen 2011).  An
interesting scenario that still needs to be investigated, and which
may be central to the present paper, is the behavior of a collimated
jet within a small residual H envelope as one might expect in SNe of
Types II-L, IIb, or IIn.

Previous evidence for asymmetry in SNe~II comes mainly from
polarization studies.  In two well-studied SNe~II-P, polarization
increased while transitioning to the nebular phase, suggesting
increasing asymmetry deeper in the expanding ejecta (Leonard et al.\
2001, 2006).  This seems consistent with increasing levels of
polarization with decreasing H envelope mass in different SN types
(Trammell et al.\ 1993; H\"oflich et al.\ 1996; Tran et al.\ 1997;
Leonard et al.\ 2000; Wang et al.\ 2001; Chornock et al.\ 2011),
whereas the massive H envelope of SN~1987A had weaker polarization
(Jeffery 1991 and references therein).  A bipolar distribution of
$^{56}$Ni was also inferred for SN~1999em and SN~2004dj based on blue
and red humps in the H$\alpha$ line profile (Elmhamdi et al.\ 2003;
Chugai et al.\ 2005); this is similar to the argument we present below
for SN~2010jp, but the blue and red humps in SN~2010jp are much
faster, more distinct, and are persistent over mutiple epochs.  

Although there is evidence for some degree of asymmetry seen in some
SNe II, there has been no claim of a clear detection of a highly
collimated bipolar jet in a SN II.  In this paper we discuss kinematic
evidence of a jet in a Type II explosion based on the unusual
triple-peaked line profiles in optical spectra of SN~2010jp.

\section{Discovery and Environment of SN~2010\lowercase{jp}
  (PTF\lowercase{aaxi})}

SN~2010jp was discovered on 2010 Nov.\ 11.3 by Maza et al.\ (2010),
and was found independently on 2010 Nov.\ 16 as PTF10aaxi in the
course of the Palomar Transient Factory (PTF; Law et al.\ 2009; Rau et
al.\ 2009), as we describe below.  Challis, Kirshner, \& Smith (2010)
obtained a spectrum that showed peculiar line profiles, with narrow
peaks in the Balmer lines resembling a Type~IIn like SN~2006gy (Smith
et al.\ 2007, 2010; Ofek et al. 2007), but also with very broad line
wings extending to at least $\pm$15,000 km s$^{-1}$ at early times.
Throughout, we adopt $E(B-V)$=0.087 mag and $A_R$=0.233 mag as the
Galactic reddening and $R$-band extinction in the direction of
SN~2010jp (Schlegel et al.\ 1998).  At $z$ = 0.009 measured from the
narrow H$\alpha$ peak in our spectra (see below), we adopt $m-M$ =
32.9 mag.  Corrected for extinction, SN~2010jp then has a peak
absolute unfiltered (taken to be approximately $R$-band) magnitude of
$-$15.9.  This is on the low-luminosity tail of the luminosity
functions for SNe~II-L and SNe~IIn (Li et al.\ 2011).  In this paper
we present additional photometric and spectroscopic observations of
SN~2010jp, and we suggest that high-speed material seen in spectra
arises in a nonrelativistic jet.
 
There is no host galaxy known at the position of SN~2010jp.
Figure~\ref{fig:img} shows an image of the surrounding field.  In our
pre-explosion PTF images taken between 2009 November and 2010 March,
we detected no underlying host galaxy emission to a limiting magnitude
of $R$$>$21.7 (see below).  This may suggest that the host is a very
faint and low-metallicity dwarf galaxy with an absolute magnitude
fainter than about $-$12, roughly 40 times fainter than the Small
Magellanic Cloud.  Such a faint host is very rare among core-collapse
SNe (Arcavi et al.\ 2010). Alternatively, the position of SN~2010jp is
seen about 3\arcmin \ in projection from the center of the pair of
interacting galaxies NGC~2207/IC~2163 (Figure~\ref{fig:img}).
NGC~2207 has a redshift of $z$=0.0091, which is consistent with that
of SN~2010jp.  If they are associated, this would put SN~2010jp at a
galactocentric radius of about 33 kpc, in the outer regions of this
system.  This is at the tail of the distribution for host-SN distances
among PTF SNe (Kasliwal et al., in prep.).  In either case, SN~2010jp
occurred in a very low-density and probably also a low-metallicity
region, although we do not have any measurement of the true
metallicity in the environment.  This remote environment may be a key
ingredient in producing such an unusual jet-driven SN.

\begin{figure}\begin{center}
\includegraphics[width=3.3in]{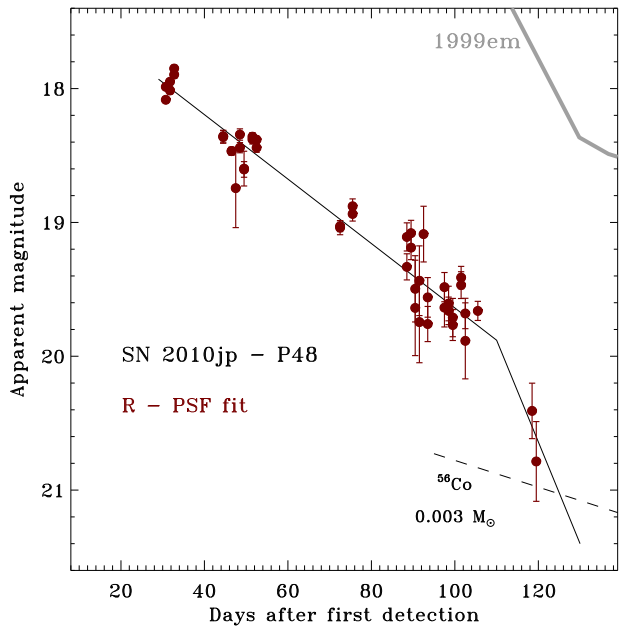}
\end{center}
\caption{P48 photometry of SN~2010jp, plotted as days after the first
  detection. We show $R$-band magnitudes from the Palomar 48-in
  telescope, reduced using PSF-fitting photometry.  The error bars are
  from photon counting statistics, but the scatter suggests that the
  true uncertainty is somewhat larger due to systematic effects
  associated with the bright nearby star.  Therefore, the solid black
  line shows an idealized decline with a fading rate of 0.024 mag
  d$^{-1}$ until day 110, and a faster fading rate of 0.076 mag
  d$^{-1}$ after day 110.  The dashed line shows the expected
  $^{56}$Co radioactive decay luminosity for M($^{56}$Ni) = 0.003
  $M_{\odot}$, and the gray line in the upper right corner is
  SN~1999em as it would look at the same distance and reddening (see
  Figure~\ref{fig:phot}).}\label{fig:photP48}
\end{figure}

\begin{figure}\begin{center}
\includegraphics[width=3.3in]{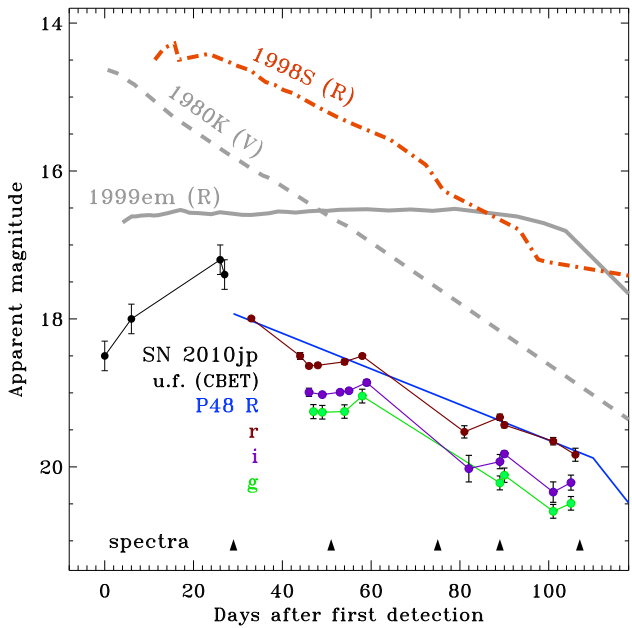}
\end{center}
\caption{Multi-filter light curve of SN~2010jp, plotted as days after
  the first detection. We show unfiltered magnitudes taken from
  preliminary reports (Maza et al.\ 2010), plus $g$, $r$, and $i$-band
  photometry from the Palomar 60-in telescope. The solid blue line
  shows the approximate decline rate of $R$-band magnitudes from the
  Palomar 48-in telescope, reproduced from
  Figure~\ref{fig:photP48}. For comparison, we also show the $R$-band
  light curves of the normal SN~II-P 1999em (Leonard et al.\ 2002;
  solid gray line) and the SN IIn 1998S (Fassia et al.\ 2000; orange
  dot-dashed line), as well as the $V$-band light curve of the SN~II-L
  1980K (Buta 1982; grey dashed line) as they would appear at the same
  presumed distance of SN~2010jp and with the same assumed foreground
  extinction. The error bars here and in the data tables reflect
  photon statistics; the uncertainty in the absolute calibration is
  somewhat larger, typically 0.15 mag, due to a nearby star.
  Arrowheads at the bottom denote dates when we obtained spectra of
  SN~2010jp.}\label{fig:phot}
\end{figure}

\begin{figure}\begin{center}
\includegraphics[width=3.3in]{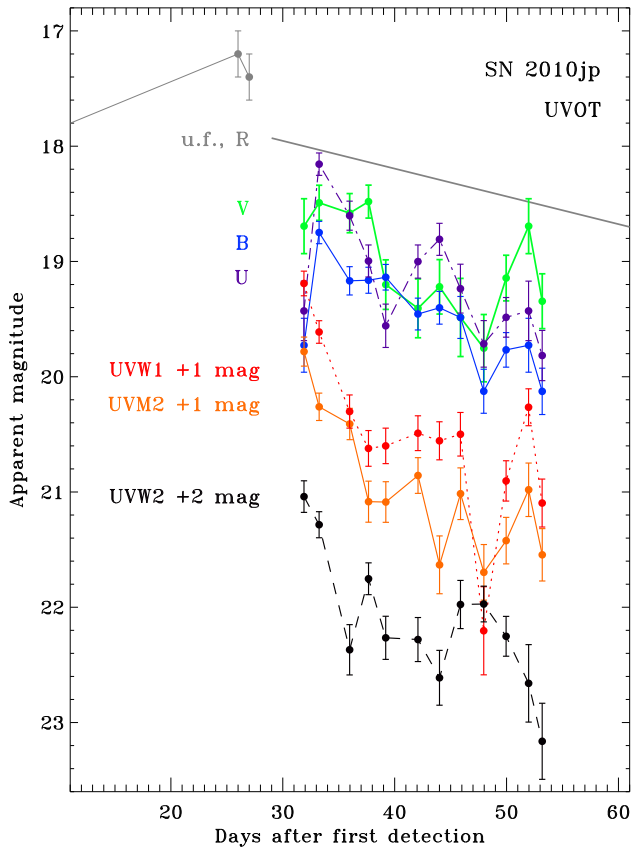}
\end{center}
\caption{Lightcurves derived from {\it Swift}/UVOT photometry in the
  $V$ (green), $B$ (blue), $U$ (purple, dot-dashed line), $UVW1$ (red,
  dotted line, +1 mag offset), $UVM2$ (orange, +1 mag offset), and
  $UVW2$ (black, dashed line, +2 mag offset) filters.  The unfiltered
  magnitudes and $R$-band decline rate from Figure~\ref{fig:phot} are
  reproduced in gray for comparison.}\label{fig:uvot}
\end{figure}

\begin{table}\begin{center}\begin{minipage}{2.5in}
      \caption{P48 $R$-band magnitudes for SN~2010jp using PSF-fitting
        photometry.}
\scriptsize
\begin{tabular}{@{}lcc}\hline\hline
JD &$R$ mag &$R$ err$^a$ \\ \hline
2455516.90    &18.08 &0.02 \\
2455516.95    &17.98 &0.02 \\
2455517.90    &18.01 &0.03 \\
2455517.95    &17.95 &0.02 \\
2455518.80    &17.90 &0.03 \\
2455530.81    &17.90 &0.03 \\
2455532.79    &18.47 &0.03 \\
2455532.83    &18.47 &0.02 \\
2455533.70    &18.74 &0.29 \\
2455534.82    &18.44 &0.04 \\
2455534.83    &18.34 &0.04 \\
2455535.79    &18.60 &0.05 \\
2455535.81    &18.59 &0.13 \\
2455537.77    &18.36 &0.03 \\
2455537.79    &18.38 &0.03 \\
2455538.77    &18.44 &0.03 \\
2455538.81    &18.38 &0.03 \\
2455558.82    &19.04 &0.05 \\
2455558.87    &19.03 &0.04 \\
2455561.76    &18.94 &0.05 \\
2455561.81    &18.88 &0.05 \\
2455574.67    &19.11 &0.11 \\
2455574.72    &19.33 &0.10 \\
2455575.69    &19.08 &0.10 \\
2455575.74    &19.19 &0.09 \\
2455576.70    &19.64 &0.36 \\
2455576.75    &19.50 &0.25 \\
2455577.71    &19.44 &0.26 \\
2455577.75    &19.74 &0.30 \\
2455578.79    &19.09 &0.21 \\
2455579.76    &19.76 &0.13 \\
2455579.81    &19.56 &0.15 \\
2455583.71    &19.48 &0.11 \\
2455583.75    &19.64 &0.14 \\
2455584.70    &19.66 &0.10 \\
2455584.75    &19.61 &0.13 \\
2455585.71    &19.77 &0.12 \\
2455585.76    &19.71 &0.14 \\
2455587.67    &19.41 &0.08 \\
2455587.72    &19.47 &0.10 \\
2455588.73    &19.68 &0.11 \\
2455588.78    &19.88 &0.28 \\
2455591.75    &19.66 &0.07 \\ 
2455604.67    &20.41 &0.21 \\
2455605.72    &20.79 &0.30 \\
\hline
\end{tabular}\label{tab:p48}\end{minipage}

$^a$Error bars refer to photon-counting statistics.  Uncertainty in
the absolute calibration is approximately 0.15 mag.
\end{center}
\end{table}

\begin{table}\begin{center}\begin{minipage}{3.2in}
      \caption{P60 $g$, $r$, and $i$-band magnitudes for SN~2010jp}
\scriptsize
\begin{tabular}{@{}lcccccc}\hline\hline
JD           &$g$ mag &$g$ err  &$r$ mag  &$r$ err  &$i$ mag &$i$ err  \\ \hline
2455518.87   &...   &...    &17.99 &0.10  &...   &...    \\
2455529.84   &...   &...    &18.50 &0.11  &...   &...    \\
2455531.83   &...   &...    &18.64 &0.11  &18.99 &0.077  \\
2455532.83   &19.25 &0.095  &...   &...   &...   &...    \\
2455533.83   &...   &...    &18.63 &0.11  &...   &...    \\
2455534.85   &19.26 &0.092  &...   &...   &19.02 &0.078  \\  
2455538.81   &...   &...    &...   &...   &18.99 &0.079  \\
2455539.81   &19.25 &0.091  &18.58 &0.11  &...   &...    \\
2455540.81   &...   &...    &...   &...   &18.97 &0.081  \\
2455543.80   &19.04 &0.093  &18.50 &0.11  &...   &...    \\
2455544.80   &...   &...    &...   &...   &18.86 &0.084  \\
2455566.74   &...   &...    &19.53 &0.11  &...   &...    \\
2455567.73   &...   &...    &...   &...   &20.02 &0.082  \\
2455574.72   &20.22 &0.094  &19.33 &0.10  &19.93 &0.080  \\
2455575.82   &20.11 &0.097  &19.43 &0.11  &19.83 &0.081  \\
2455586.76   &20.60 &0.093  &19.65 &0.11  &20.34 &0.082  \\
2455590.67   &20.49 &0.092  &...   &...   &20.21 &0.079  \\
2455591.67   &...   &...    &19.84 &0.11  &...   &...    \\
\hline
\end{tabular}\label{tab:p60}
\end{minipage}\end{center}
\end{table}

\section{OBSERVATIONS}

\subsection{Palomar 48-inch discovery and photometry}

We obtained $R$-band images of PTF field 100080 on UT 2010 Nov.\ 16
(UT dates are used throughout) with the Palomar 48-inch telescope
(P48) equipped with the refurbished CFHT12k camera (Rahmer et
al. 2008). Subtraction of a stacked reference image of the field with
HOTPANTS\footnote{http://www.astro.washington.edu/users/becker/hotpants.htm}
revealed a new transient source at coordinates $\alpha$ =
06$^{\mathrm{h}}$16$^{\mathrm{m}}$30$\fs$63, $\delta$ =
$-21\fd$24$\farcm$36$\farcs$2 (J2000.0), with an astrometric
uncertainty (relative to the USNO-B1 catalog; Monet et al.\ 2003) of
$\pm$150\,mas in each coordinate.  The transient was discovered $30$
hours later by Oarical, an autonomous software framework of the PTF
collaboration (Bloom et al.\ 2011b).  It was classified correctly as a
transient source (as opposed to a vairable star), was further
classified as a SN or nova, and was given the name PTF10aaxi.  This
source discovered independently by PTF is the same as SN~2010jp.

No source was detected at this location with P48 in a combined image
taken from 2009 November to 2010 March, 8--12 months prior to
discovery, to a 3-$\sigma$ limiting magnitude of $R$ $\approx$ 21.7
(see Figure~\ref{fig:img}).  The limiting magnitude in the co-added
image is actually $R$ $\approx$ 22.3 mag, but SN~2010jl is located
only a few arcseconds from a bright foreground star that reduces the
effective sensitivity of the image at that location.  

A listing of P48 observations taken around the time of outburst,
calibrated to USNO-B1 $R$-band (i.e., Vega-based) magnitudes of nearby
point sources is provided in Table~\ref{tab:p48}.  Because of
complications introduced by the nearby star, we analyzed the
photometry of SN~2010jp using a PSF-fitting routine for the
photometry.  In each image frame, the PSF is determined from nearby
field stars, and this average PSF is then fit at the position of the
SN event weighting each pixel according to Poisson statistics,
yielding a SN flux and flux error.  The resulting magnitudes are given
in Table~\ref{tab:p48} and are shown in Figure~\ref{fig:photP48}.  We
find that photon counting statistics underestimate the uncertainty
that is indicated by the observed scatter.  We therefore place little
emphasis on the minor undulations in the $R$-band light curve, and
instead adopt a representative smooth decline rate shown by the solid
line in Figure~\ref{fig:photP48}.  The last two measurements in
Figure~\ref{fig:photP48} were detected with rather large uncertainty.
If these late detections are reliable, they imply a somewhat faster
rate of fading after day 110.

\subsection{Palomar 60-inch photometry}

Upon discovery of PTF10aaxi, the field was automatically inserted into
the queue of the robotic Palomar 60-inch telescope (P60; Cenko et al.\
2006) for multi-colour follow-up observations (Gal-Yam et al.\
2011). Images were processed using our custom realtime pipeline, and
photometry was performed using a point-spread-function (PSF) matching
technique to remove any contribution from the nearby star USNOB-1
0685-0078200.  The $g^{\prime}$, $r^{\prime}$, and $i^{\prime}$-band
images were calibrated to USNO-B1 with filter transformations from
Jordi et al.\ (2006), so they are on the SDSS/AB photometric system
(Oke \& Gunn 1983). A log of our P60 observations of PTF10aaxi is
provided in Table~\ref{tab:p60}.  The light curve, combining the P48
and P60 photometry, is shown in Figure~\ref{fig:phot}.  We also
performed an independent analysis of the P60 photometry with the {\tt
  mkdifflc} routine (Gal-Yam et al.\ 2004). The routine uses image
subtraction with the common point spread function (PSF) method (CPM,
Gal-Yam et al.\ 2008) for PSF matching. We inserted ``artificial'' SNe
at a similar brightness than that of the real SN, using their
post-subtraction magnitude scatter as an estimate of the error due to
subtraction residuals. The results (not shown) were comparable to the
first analysis of P60 data, although there were differences that
exceeded the sizes of error bars.  As with the P48 photometry, the
nearby star probably increased the observed scatter, so we do not
emphasize small scale variation in the P60 light curves.  The general
rate of fading in P60 $gri$ data is consistent with the P48 $R$-band
data.

\begin{figure*}\begin{center}
\includegraphics[width=6.3in]{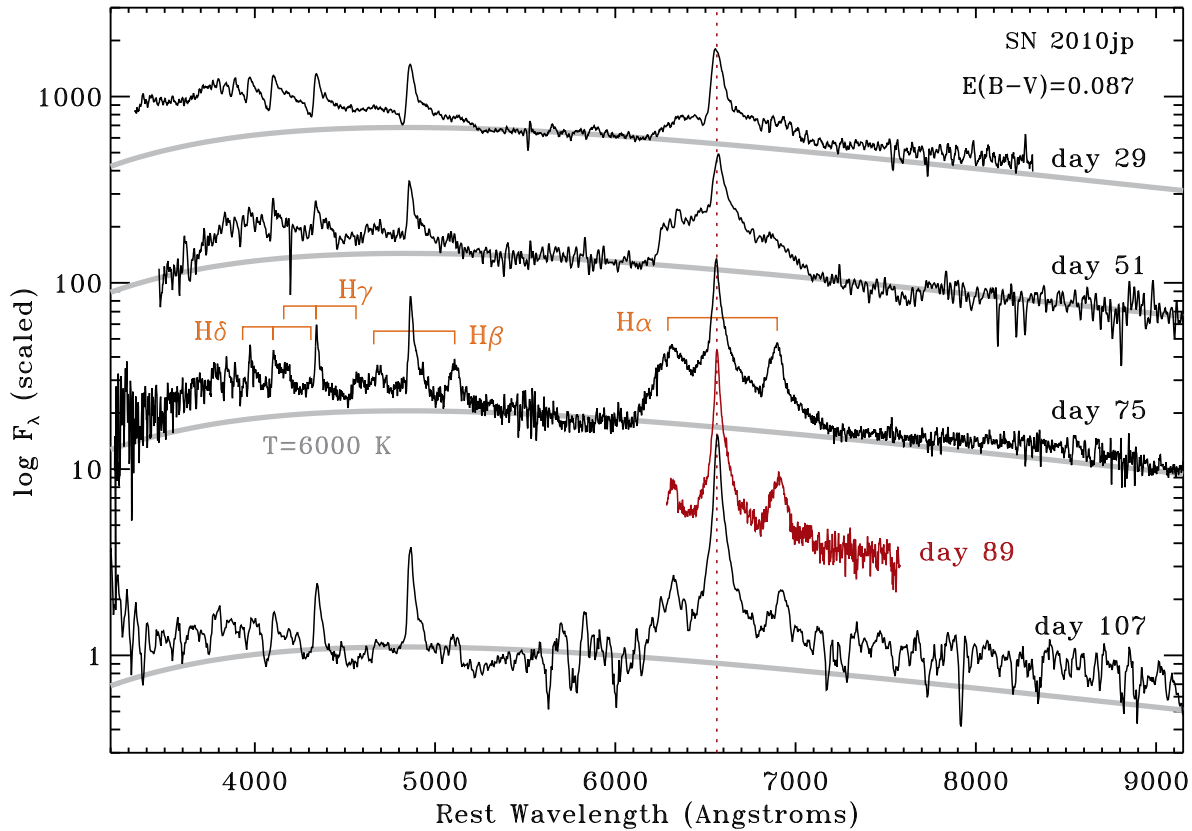}
\end{center}
\caption{Visual-wavelength spectra of SN~2010jp obtained with the MMT,
  the Palomar 200-in telescope, and Keck (see
  Table~\ref{tab:spectab}), listed as days after discovery.  The day
  89 spectrum plotted in red is a high-resolution spectrum from the
  MMT. The blue and red high-velocity components of H$\alpha$,
  H$\beta$, H$\gamma$, and H$\delta$ are marked in orange brackets on
  the day 75 spectrum. }\label{fig:spec}
\end{figure*}

\subsection{\textit{Swift} UV and X-ray Observations}

The field of SN\,2010jp was observed by the UltraViolet Optical
Telescope (UVOT; Roming et al.\ 2005) on the \textit{Swift} satellite
(Gehrels et al.\ 2004) beginning at 17:01 UT on 2010 November 17.  We
downloaded the UVOT images from the NASA HEASARC archive\footnote{See
  http://heasarc.gsfc.nasa.gov.}, and photometered the images
following the recipes provided by Li et al.\ (2006) ($V$, $B$, and
$U$-bands) and Poole et al.\ (2008) ($UVW1$, $UVM2$, and $UVW2$
filters).  The resulting measurements are plotted in
Figure~\ref{fig:uvot}.

Simultaneously, the location of SN\,2010jp was observed with the
\textit{Swift} X-ray Telescope (XRT; Burrows et al.\ 2005).  We
downloaded the XRT observations from the HEASARC archive and reduced
individual orbits following the techniques described by Butler (2007).
Stacking all available X-ray observations of this location (a total of
28.5\,ks of exposure time), no source is detected at the location of
SN\,2010jp.  Assuming a photon spectral index of $\Gamma = 2$, we
infer an upper limit on the 0.3--10\,keV X-ray flux of $< 1.5 \times
10^{-14}$\,erg\,cm$^{-2}$\,s$^{-1}$, or $<$ 5.0$\times$10$^{-4}$
counts \ s$^{-1}$.

\subsection{EVLA Observations}

We observed the field of SN\,2010jp using the Expanded Very Large
Array (EVLA; Perley et al.\ 2009) on two separate occasions: 2011
January 11 (C configuration) and 2011 May 12 (BnA configuration).
This was part of an observing program entitled ``Transients in the
Local Universe'' (P.I., M.\ Kasliwal). Both observations were
performed in the X-band (8.46 Ghz) with a total integration time of
14.0 and 15.4 min on source, respectively, and a total bandwidth of
256 Mhz. The EVLA data were reduced using the Astronomical Image
Processing System (AIPS) software.\footnote{http://www.aips.nrao.edu}
\ The calibration was done using 3C138 as the primary flux calibrator
and J0609-1542 as the phase calibrator. A final image was produced
using the AIPS IMAGR task with a pixel size of 0\farcs15. There is no
detection at the source position to a 3-$\sigma$ flux limit of
$f_{\nu} < 48$\,$\mu$Jy (Jan. 11) and $f_{\nu} < 120$\,$\mu$Jy (May
12).

\begin{table}\begin{center}\begin{minipage}{3.25in}
      \caption{Spectroscopic observations of SN~2010jp}
\scriptsize
\begin{tabular}{@{}llcccc}\hline\hline
  Date &Tel./Inst. &Day &$\delta\lambda$ (\AA) &$\lambda$/$\Delta\lambda$ &$W_{H\alpha}$(\AA) \\ 
  \hline
2010\,Nov.\,14   &MMT/B.C.  &29  &3176--8390  &1700      &405   \\
2010\,Dec.\,06   &Pal5/DBSP &51  &3500--10000 & 800/500  &777   \\
2010\,Dec.\,30   &Keck/LRIS &75  &3251--10180 & 600/1100 &1305  \\
2011\,Jan.\,13   &MMT/B.C.  &89  &6350--7650  & 4500     &...   \\
2011\,Jan.\,31   &Keck/LRIS &107 &3100--10220 & 600/1100 &1960  \\
\hline
\end{tabular}\label{tab:spectab}
\end{minipage}\end{center}
\end{table}

\subsection{Spectroscopy}

\begin{figure}\begin{center}
\includegraphics[width=3.2in]{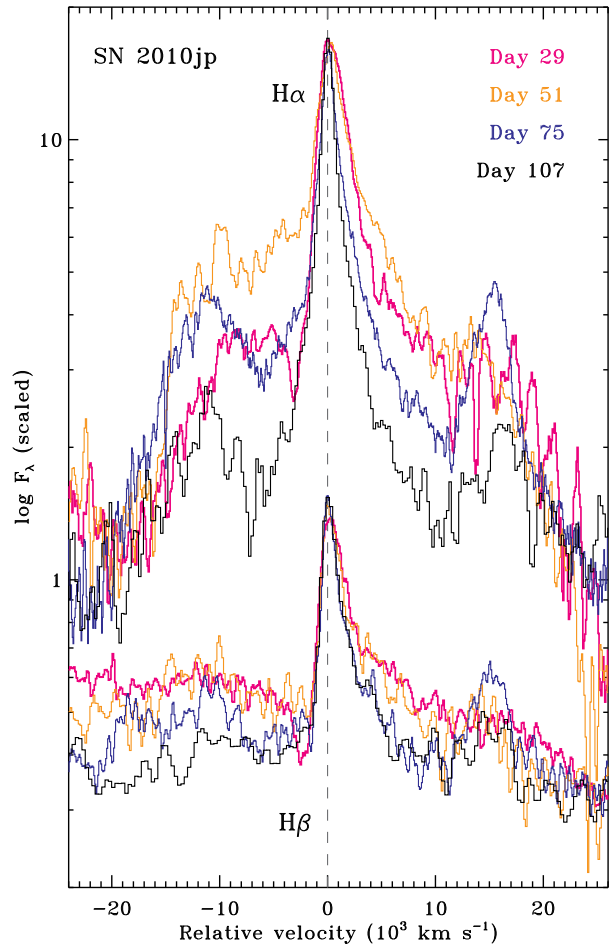}
\end{center}
\caption{Evolution of the H$\alpha$ and H$\beta$ profiles of
  SN~2010jp.}\label{fig:halpha}
\end{figure}

\begin{figure}\begin{center}
\includegraphics[width=3.2in]{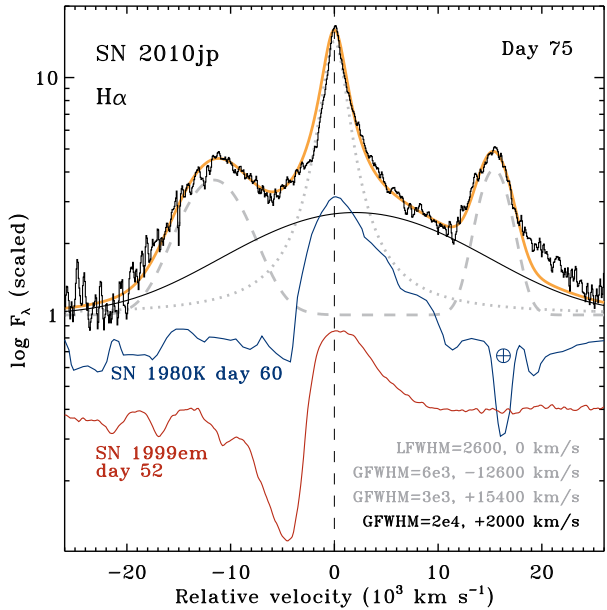}
\end{center}
\caption{The H$\alpha$ profile of SN~2010jp on day 75, decomposed into
  multiple contributing features (thin black and grey curves) with the
  sum of all individual components shown in orange. GFWHM and LFWHM
  denote Gaussian or Lorentzian FWHM values and centroid velocities.
  For comparison, we also show the H$\alpha$ profile of the normal SN
  II-P 1999em from Leonard et al.\ (2002), plotted in red, as well as
  the H$\alpha$ line in the Type II-L SN~1980K observed on day
  $\sim$60 (Barbieri et al.\ 1982) in blue (the absorption feature
  marked $\earth$ in the SN~1980K spectrum is an uncorrected telluric
  feature).}\label{fig:halpha2}
\end{figure}

After discovery, we obtained several epochs of optical spectroscopy of
SN~2010jp using the Bluechannel spectrograph on the 6.5-m Multiple
Mirror Telescope (MMT), the Double Beam Spectrograph (DBSP; Oke \&
Gunn 1982) on the Palomar 200-in telescope (P200), and the
Low-Resolution Imaging Spectrometer (LRIS; Oke et al.\ 1995) mounted
on the 10-m Keck~I telescope.  Details of the spectral observations
are summarized in Table~\ref{tab:spectab}.  The slit was always
oriented at the parallactic angle (Filippenko 1982), and the long-slit
spectra were reduced using standard procedures.  Final spectra are
shown in Figure~\ref{fig:spec}.  Details of the H$\alpha$ line profile
are shown in Figures~\ref{fig:halpha} and \ref{fig:halpha2}.

\begin{figure}\begin{center}
\includegraphics[width=3.2in]{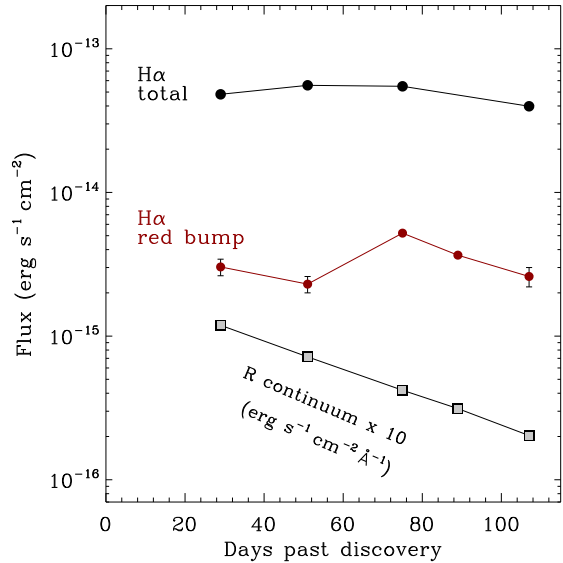}
\end{center}
\caption{The total H$\alpha$ line flux (black dots) and the flux
  contained in just the red high-velocity component (red dots), as
  compared to the decline in the continuum flux over time (grey
  squares; flux density multiplied by a factor of 10 for display
  here).  The error bars for the red component flux are shown, and the
  error bars for the total H$\alpha$ flux are smaller than the
  plotting symbols.  These are measurement errors; there is also a
  $\sim$5\% uncertainty in the red continuum flux based on the smooth
  fit to the $r$ and $R$-band magnitudes.}\label{fig:halpha3}
\end{figure}

Table~\ref{tab:spectab} also lists approximate emission equivalent
widths (positive values for emission lines) of the H$\alpha$ line at
each epoch ($W_{H\alpha}$).  The uncertainty here is a few to 5\%,
dominated by the uncertainty in the continuum level (we could not
measure an equivalent width for the 2011 Jan.\ 13 spectrum from the
MMT, since this high-resolution spectrum did not include the continuum
on the blue side of the very broad line).  Adopting a single value for
the red continuum at each spectral epoch by fitting a straight line to
the declining $r$ and $R$ magnitudes in Figure~\ref{fig:phot}, we then
converted these equivalent widths to total fluxes in the H$\alpha$
line.  Figure~\ref{fig:halpha3} compares the decline rate of the
continuum to the total H$\alpha$ line flux, as well as the flux of the
high-velocity red bump component measured in the same spectra (we
could measure the red component flux in the MMT spectrum).  The total
H$\alpha$ line flux declines more slowly with time compared to the red
continuum, so that the equivalent width increases, as is typical for
SNe~II.  However, from Figure~\ref{fig:halpha3} we also see that the
high-velocity red bump increases its relative strength compared to the
total line flux and continuum after day $\sim$60.

\section{Analysis}

\subsection{Light Curve and Luminosity}

After reaching its peak luminosity at 20--30 days, SN~2010jp has a
linear decline rate in visual-wavelength bands that is similar to SNe
II-L and IIn, and does not show a plateau akin to SNe II-P
(Figure~\ref{fig:phot}).  There is, however, tentative evidence for an
increased fading rate after day 110, which is similar to some SNe IIn,
such as SN~1994W (Sollerman et al.\ 1998). Linear fading is not
unusual in the light curves of moderate-luminosity SNe~IIn.  A well
studied example is SNe~1998S (Fassia et al.\ 2000), shown in
Figure~\ref{fig:phot} for comparison.  Other examples of
moderate-luminosity SNe~IIn with linear declines are SN~2005gl
(Gal-Yam, et al.\ 2007), SN~2005ip (Smith et al.\ 2009b), and PTF09uj
(Ofek et al.\ 2010).  In general, the UVOT filters
(Figure~\ref{fig:uvot}) show a somewhat faster decline rate than the
$R$-band, which is typical for blue/UV decline rates of Type II SNe.
This implies that the UV filters are dominated by continuum emission
from the photosphere in the SN ejecta, and not CSM interaction.

SN~2010jp is sub-luminous compared to most SNe~II-P, II-L, and IIn.
While its overall light curve shape is quite similar to SN~1998S
(Figure~\ref{fig:phot}), it is about 2.5 mag fainter in terms of its
absolute magnitude.  It has a peak absolute magnitude of about $-$15.9
(unfiltered, approximately $R$-band), lying on the faint tail of the
luminosity function for SNe~IIn (Li et al.\ 2011).  Note that in
Figure~\ref{fig:phot}, the apparent magnitudes of SN~1999em are shown
as they would appear at the same distance and with the same reddening
as SN~2010jp.  In our spectra, we do not detect a strong narrow
absorption component in Na~{\sc i} D, but Poznanski et al.\ (2011)
find that the strength of Na~{\sc i} D is not a good proxy for
line-of-sight extinction anyway.  The characteristic visual-wavelength
continuum temperature (after correcting for Galactic reddening) of
$\sim$6000 K is consistent with other SNe~IIn, giving us no reason to
suspect a value for the local reddening much larger than about $A_V
\simeq$ 0.5 mag.  Hence, we find it unlikely that the peak of
SN~2010jp is more luminous than about $-$16 mag.

Since the peak luminosity is so low, we conjecture that CSM
interaction does not provide a substantial luminosity boost for
SN\,2010jp.  As such, the CSM interaction in SN~2010jp is similar to
SNe~IIn where the CSM interaction is strong enough to produce bright
narrow emission lines in the spectrum, but where the CSM density is
not high enough to provide efficient conversion of SN ejecta kinetic
energy into continuum radiation (see Smith et al.\ 2009a).  This
implies that the progenitor star's mass-loss rate was not more than
about 10$^{-3}$ $M_{\odot}$ yr$^{-1}$ (Smith et al.\ 2009a).  However,
at late times (50-100 days) when SN~2010jp has faded from its peak
luminosity, some of the variations in the light curve, if real, may be
due to changes in the strength of CSM interaction as the shock
overtakes density fluctuations in the CSM.  Such changes may provide
fluctuating contributions of UV continuum and line emission as the SN
photosphere fades.

SN2010jp has a relatively low peak luminosity, but its low late-time
luminosity is noteworthy as well.  The luminosity indicated by the
last two points after day 110 in Figure~\ref{fig:photP48} would be
consistent with a radioactive decay tail that is at least 10 times
less luminous than that of SN~1999em.  Adopting M($^{56}$Ni) = 0.03
$M_{\odot}$ for SN~1999em (0.02~$M_{\odot}$ found by Elmhamdi et al.\
2003, or 0.036~$M_{\odot}$ from Utrobin 2007), this would imply an
initial nickel mass less than about M($^{56}$Ni) \, $\la$ \, 0.003 \,
$M_{\odot}$ in SN~2010jp (see Figure~\ref{fig:photP48}), if increased
extinction from newly formed dust can be neglected.  This is a very
low nickel mass for a core collapse SN, comparable to that inferred
for SN~1994W (although SN~1994W had a much higher peak luminosity;
Sollerman et al.\ 1998).  It could indicate either that the progenitor
star had a relatively low initial mass near 8 $M_{\odot}$, or that it
was a substantially more massive star whose SN was underluminous
because it lost much of the radioactive material into a black hole
(see below).

\subsection{Spectra}

While the light curve for SN~2010jp is not very remarkable, its
spectrum is quite unusual.  The spectrum has a blue continuum with a
characteristic temperature close to $\sim$6,000 K, and strong Balmer
lines that have narrow cores and broad wings (Figure~\ref{fig:spec}).
These properties are typical of some SNe~IIn, especially those like
SN~1988Z and SN~2005ip, where both a broad and narrow H$\alpha$
component can be seen clearly (Chugai \& Danziger 1994; Smith et al.\
2009b).  In these SNe, the narrow Balmer emission is thought to arise
from dense pre-shock circumstellar material (CSM) or material in a
cold dense shell that has been hit by the forward shock, whereas the
broader line components arise either in the freely expanding SN ejecta
or the SN ejecta that have passed the reverse shock (see Chugai \&
Danziger 1994; Smith et al.\ 2008a, 2009b, 2010).  Note, however, that
broad line wings of a few 10$^3$ km s$^{-1}$ in SNe~IIn can also arise
from electron-scattering wings if they have the characteristic smooth
Lorentzian profile (see Chugai 2001; Dessart et al.\ 2009; Smith et
al.\ 2010). Thus, the most basic of SN~2010jp's observed properties --
the light curve shape and overall character of the spectrum -- seem
well-explained by a Type IIn supernova.  

In our high-resolution MMT spectrum taken on day 89, the narrowest
emission component of H$\alpha$ is fully resolved, with a Gaussian
FWHM no narrower than about 800 km s$^{-1}$.  This indicates
circumstellar material that is considerably faster than the 100-200 km
s$^{-1}$ speeds commonly observed in the CSM of some SNe~IIn, but in
line with the SNe~IIn CSM velocities in the sample studied by Kiewe et
al.\ (2010).  Expansion speeds around 800 km s$^{-1}$ are much faster
than can be achieved in a red supergiant wind, but such a speed is
typical for the line widths observed in luminus blue variable
(LBV)-like eruptions (Smith et al.\ 2011).

The key spectral feature of SN~2010jp that stands out compared to all
other known SNe is that H$\alpha$ and other Balmer lines show fast
blue and red emission bumps, yielding a pronounced {\it triple-peaked}
line profile (see Figures~\ref{fig:spec} and \ref{fig:halpha}).  These
triple peaks persist across multiple epochs during the first 100
days after explosion.  They can be seen in higher Balmer lines like
H$\beta$ (Figure~\ref{fig:halpha}), as well as H$\gamma$ and H$\delta$
(Figure~\ref{fig:spec}), so the blue and red humps are definitely not
contamination by unidentified emission lines on either side of
H$\alpha$.  These blue and red emission bumps are strong enough that
in the higher order Balmer lines, they blend together to form an
excess blue pseudo-continuum (Figure~\ref{fig:spec}).
Figure~\ref{fig:halpha2} breaks down the line profile into a narrow
Lorentzian profile, a very broad Gaussian, and two well-defined
Gaussian bumps centered at roughly $-$13,000 and $+$15,000 km
s$^{-1}$, as indicated by our high signal-to-noise ratio spectrum on
day 75.  This is only meant to demonstrate that the blue and red humps
are distinct, separate emission components.  These bumps seem to shift
slightly in velocity from one epoch to the next, but they do not stray
far from those fiducial velocities, and the low signal to noise of
some of the spectra make precise velocities difficult to measure.
They do not, however, exhibit a systematic migration in velocity with
time.

Figure~\ref{fig:halpha2} shows that the H$\alpha$ line wings and the
blue and red emission humps in SN~2010jp are at very high speeds
compared to those of prototypical SNe II-L and II-P (i.e. SNe~1980K
and 1999em, respectively) at comparable epochs after explosion.  The
blue and red humps are also well separated from the central
narrow/intermediate-width emission component, which by itself would
not have been unusual, and the relative flux of the high-velocity
components appears to fade at a rate that is different from the rest
of the line (Figure~\ref{fig:halpha3}; see below).  This slow fading
implies densities of $n_H \la 10^6$ cm$^{-3}$ for the H$\alpha$
emitting region based on the expected cooling timescale.

At no time does the spectrum of SN~2010jp exhibit emission or
absorption features associated with Fe~{\sc ii}, and we do not see the
IR Ca~{\sc ii} triplet at late times that is usually seen in SNe II.
Assuming that these features com from Ca and Fe that was present in
the progenitor star's envelope, this may be another indication of very
low progenitor metallicity, as inferred from the SN host environment
(see above).

\begin{figure}\begin{center}
\includegraphics[width=3.2in]{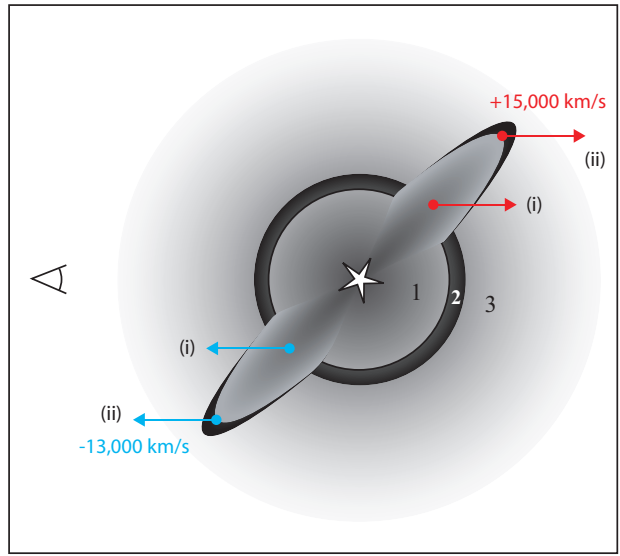}
\end{center}
\caption{Cartoon of the possible jet-powered geometry in SN~2010jp.
  Regions 1, 2, and 3 correspond to the unshocked SN ejecta (inner
  gradient), the CSM interaction region (dark), and the pre-shock CSM
  (outer gradient), respectively.  The upper-left and lower-right
  quadrants in this cartoon are the same as for a conventional SN~IIn
  with CSM interaction, corresponding to low/mid-latitudes in the
  explosion.  The lower-left and upper-right quadrants depict a tilted
  fast bipolar jet breaking through the otherwise spherical CSM
  interaction shell.  An observer located to the left would see a
  combination of the spectrum from a conventional SN~IIn, plus blue
  and red emission peaks in emission lines arising in either the
  unshocked jet material (i) or at the reverse shock in the jet (ii).
  We suggest that a bipolar jet such as this causes the blue and red
  bumps at $-$13,000 and $+$15,000 km s$^{-1}$ observed in the
  H$\alpha$ profile of SN~2010jp.}\label{fig:cartoon}
\end{figure}

\section{Discussion}

We speculate that the observed properties of SN~2010jp can arise from
a superposition of two physical scenarios.  The first is a relatively
traditional Type~IIn explosion, where the rapidly expanding low-mass H
envelope of the star collides with dense pre-existing CSM that was
ejected recently by the progenitor.  This can produce the blue
continuum, the narrow emission cores of the Balmer lines, and some of
the underlying broad emission profiles the same way that these arise
in traditional SNe IIn (see, e.g., Chugai \& Danziger 1994; Smith et
al.\ 2008a, 2009b, 2010; Kiewe et al.\ 2010).  We suspect that most of
the emitting volume in SN~2010jp corresponds to this component
(regions 1, 2, and 3 in Figure~\ref{fig:cartoon}).

The second component superposed on this model is that SN~2010jp also
produces a fast bipolar jet, tilted out of the plane of the sky, which
gives rise to the fast blue and red emission features in H$\alpha$,
and may produce some of the emission in the very broad wings of the
line.  The combination of these two scenarios is depicted
schematically in Figure~\ref{fig:cartoon}.

We argue that the two isolated red and blue emitting components must
arise in a collimated geometry.  If the fast and slow components arose
in a spherical geometry (i.e., a fast blast wave overrunning slow
clumps; Chugai \& Danziger 1994; Smith et al.\ 2009b), this would
yield both broad and narrow components, but the fast material would
form a broad component distributed over all velocities, not in blue
and red peaks.  Similarly, the reverse shock of fast ejecta plowing
into a dense slow equatorial ring (as seen presently in SN1987A;
Michael et al.\ 2003; Smith et al.\ 2005) would produce emission
spread over all velocities if it is spatially unresolved.  Two
distinct, nearly symmetric blue and red components argue for a jet,
because this can produce fast emitting material at very specific
velocities, separated from the rest of the emitting ejecta
(Figure~\ref{fig:cartoon}).  In hindsight, it would obviously have
been useful to obtain spectrapolarimetry of SN~2020jp to put
additional constraints on the asymmetry (Wang \& Wheeler 2008), but
unfortunately we did not obtain these data.

The detection of a collimated jet in a Type II explosion is
unprecedented.  Models of jet-powered SNe have been published for
fully stripped-envelope progenitors, which yield SNe~Ibc and GRBs.
Theoretical models also predict jet-driven SNe~II for a wide range of
initial masses exceeding 25 $M_{\odot}$ (e.g., Heger et al.\ 2003),
from collapsars that yield black holes.  These are expected to be more
common at sub-solar metallicity (Heger et al.\ 2003), due to the
expectation of weaker metallicity-dependent mass loss.  However, no
clear case of a jet-driven SN~II has yet been seen.  For normal red
supergiants (RSGs) with massive H envelopes, one expects that the
collimated jet is largely destroyed while imparting its kinetic energy
to a spherical envelope (MacFadyan et al.\ 2001; H\"oflich et al.\
2001; Wheeler et al.\ 2002; Couch et al.\ 2009).  It would be
interesting to conduct similar hydrodynamic and radiative transfer
simulations for a jet-driven SN from a progenitor that had a very
small residual H envelope, perhaps due to the same type of episodic
pre-SN mass loss that leads to a Type IIn event.  This type of
episodic mass loss may be driven by continuum radiation opacity or may
be hydrodynamic, and is therefore not necessarily sensitive to
metallicity (Smith \& Owocki 2006). For a very small amount of H
remaining on the star's surface, the jet might survive traversal of
the envelope.

We consider two possible specific mechanisms that may power the
emission in the fast blue and red peaks of the H Balmer series:

(i) A jet-powered explosion might mix significant quantities of
$^{56}$Ni to high velocities in the polar regions of a thin H
envelope. The radioactivity would then heat that H-rich ejecta
directly at those high velocities.  In fact, it is interesting that
the H$\alpha$ profile we observe in SN~2010jp bears a striking
resemblence to the distribution of $^{56}$Ni velocities in simulations
of jet-powered SNe (Couch et al.\ 2009; see their Figure 12).  A
similar mechanism was proposed to explain similar (although much
weaker) features in the H$\alpha$ profiles of the SNe II-P 1999em and
2004dj (Chugai et al.\ 2005; Elmhamdi et al.\ 2003).  The lack of
prominent Fe lines in the late-time spectrum may be a concern for this
hypothesis, however.

(ii) Alternatively, without $^{56}$Ni mixed to high velocities, the
fast blue and red H$\alpha$ bumps may be excited by CSM interaction.
Fast H-bearing SN ejecta that cross the reverse shock of the jet will
also yield two very fast but localized velocity components
(Figure~\ref{fig:cartoon}).  This option is difficult to rule out,
given the evidence for CSM interaction in the Type~IIn spectrum.

Choosing between these two will require additional theoretical work,
including hydrodynamic radiative transfer simulations of jets in a
light H envelope.  Both of these require some H to be present at high
velocities in the polar regions of the SN ejecta; therefore, a model
wherein H only exists in the CSM (i.e., with a true stripped-envelope
SN~Ibc and jet plowing into dense H-rich CSM) cannot explain the
observations of SN~2010jp.  This may present an interesting challenge
to current models of jets in core-collapse SNe.

Examining Figure~\ref{fig:halpha3}, we see that at times after 50--60
days, the strength of the red high-velocity emission component (and
also the blue component; not shown) appears to increase relative to
that of the total H$\alpha$ line flux of the continuum flux density.
This may mark the time when the recombination photosphere recedes
through the faster parts of the ejecta to reveal the jet material
heated radioactively in scenario (i), or when the fast jet overtakes
and breaks through the optically thick CSM interaction shell in
scenario (ii).  The size of the $\sim$6000\,K continuum photosphere at
this time should be very roughly 30~AU.

From our observations, it is not obvious if the jet in SN~2010jp was
driven by accretion onto a newly born black hole or by a magnetar jet.
Models of magnetar outflows colliding with SN ejecta have been invoked
as a possible origin for some of the most luminous SNe known (Kasen \&
Bildsten 2010; Woosley 2010), but in SN~2010jp we see evidence for a
collimated jet in a relatively {\it faint} SN with only moderate CSM
interaction.  We showed earlier that the peak absolute magnitude of
SN~2010jp was 1--3 mag less luminous than typical SNe~II-L and IIn,
and that the late-time luminosity suggests a very small initial nickel
mass of M($^{56}$Ni) \, $\la$ \, 0.003 $M_{\odot}$.  One might
speculate that a weak explosion or small yield of $^{56}$Ni might be
the consequence of fallback into a black hole in a Type II collapsar,
where some material from the accretion disk is launched poleward to
produce the jets we observe.  Jet-driven SNe II from collapsars can be
more energetic than normal SNe, but not necessarily (MacFadyan et al.\
2001; Heger et al.\ 2003).

The radio and X-ray non-detections we report for SN\,2010jp are,
unfortunately, not very constraining in terms of choosing between
options (i) and (ii) above.  Type~IIn SNe can have strong radio and
X-ray emission from CSM interaction at late times, as in the case of
SN\,1988Z (Van Dyk et al.\ 1993; Fabian \& Terlevich 1996; Schlegel \&
Petre 2006), but the radio and X-ray emission from CSM interaction can
also be quashed due to high optical depths, as in the case of
SN\,2006gy (Smith et al.\ 2008b, 2010; Ofek et al.\ 2007).  From the
radii and density quoted above for SN~2010jp, the 1 GHz free-free
continuum optical depth would be about 10$^{20}$.

Lastly, we consider the possibility that SN~2010jp was due to a
transient accretion event where the fast blue and red emission bumps
in the H$\alpha$ profile could be produced by a rotating disk around a
massive black hole, akin to the class of active galactic nuclei that
are double-peaked emitters (Halpern \& Filippenko 1988).  In order for
the emitting radius of this fast orbiting material to be larger than
the continuum photospheric radius required for SN~2010jp (about 20--60
AU at various times as the SN fades), the black hole mass would need
to be more than 5$\times$10$^6$ $M_{\odot}$.  Such an event might
resemble Sw 1644+57; Bloom et al.\ (2011a) proposed this to be a tidal
disruption event around a 10$^6$--10$^7$ $M_{\odot}$ black hole,
whereas Quataert \& Kasen (2011) discussed an alternative involving a
weak SN/GRB event.  Sw 1644+57 is coincident with the light centroid
of its host galaxy, consistent with the possibility of a massive black
hole.  However, SN~2010jp's location far outside a galactic nucleus
and the non-detection of X-ray or radio emission make the tidal
disruption hypothesis seem unlikely in this case.  In addition, we
should expect both the velocity of the blue and red peaks and the
continuum temperature to increase with time in this scenario, as the
photosphere recedes to reveal faster and hotter material in an
accretion disk around the massive black hole.  This behavior is not
observed in SN~2010jp.

\smallskip\smallskip\smallskip\smallskip
\noindent {\bf ACKNOWLEDGMENTS}
\smallskip
\footnotesize

We thank P.\ Challis and R.\ Kirshner for assistance with the MMT
observations on 2010 November, and for providing the reduced spectrum
from that night.  S.B.C. acknowledges generous financial assistance
from Gary \& Cynthia Bengier, the Richard \& Rhoda Goldman Fund,
NASA/{\it Swift} grants NNX10AI21G and GO-7100028, the TABASGO
Foundation, and NSF grant AST-0908886.  D.P.\ and E.O.O.\ are
supported by Einstein fellowships from NASA. The National Energy
Research Scientific Computing Center, which is supported by the Office
of Science of the U.S.\ Department of Energy under Contract
No. DE-AC02-05CH11231, provided staff, computational resources, and
data storage for this project. The Weizmann PTF partnership is
supported in part by grants from the Israeli Science Foundation (ISF)
to A.G. Collborative Caltech-WIS work is supported by a grant from the
Binational Science Foundation (BSF) to A.G. and S.R.K. The work of
A.G. is further supported by an FP7 Marie Curie IRG Fellowship and the
Benoziyo Center for Astrophysics, and by the Lord Sieff of Brimpton
Memorial Fund.  J.S.B.\ was partially supported by a grant from the
National Science Foundation (NSF-CDI \# 0941742).  Some of the data
presented herein were obtained at the W.M.\ Keck Observatory, which is
operated as a scientific partnership among the California Institute of
Technology, the University of California and the National Aeronautics
and Space Administration. The Observatory was made possible by the
generous financial support of the W.M.\ Keck Foundation.  The authors
wish to recognize and acknowledge the very significant cultural role
and reverence that the summit of Mauna Kea has always had within the
indigenous Hawaiian community.  We are most fortunate to have the
opportunity to conduct observations from this mountain.  The National
Radio Astronomy Observatory is a facility of the National Science
Foundation operated under cooperative agreement by Associated
Universities, Inc.


\end{document}